%% file: template.tex
\documentclass{article}

\usepackage{arxiv}

\usepackage[utf8]{inputenc} 
\usepackage[T1]{fontenc}    
\usepackage{hyperref}       
\usepackage{url}            
\usepackage{booktabs}       
\usepackage{amsfonts}       
\usepackage{amsmath}  
\usepackage{nicefrac}       
\usepackage{microtype}      
\usepackage{lipsum}
\usepackage{graphicx}
\usepackage{subcaption}

\title{Querying GI Endoscopy Images: A VQA Approach\thanks{Submitted to ImageCLEFmedical at CLEF 2025}}

\author{
 Gaurav Parajuli \\
  Johannes Kepler University Linz\\
  Linz, Austria\\
  \texttt{k12455655@students.jku.at} \\
  \texttt{parajuligaurav007@gmail.com}
}

\begin{document}
\maketitle

\begin{abstract}
VQA (Visual Question Answering) combines Natural Language Processing (NLP) with image understanding to answer questions about a given image. It has enormous potential for the development of medical diagnostic AI systems. Such a system can help clinicians diagnose gastro-intestinal (GI) diseases accurately and efficiently. Although many of the multimodal LLMs available today have excellent VQA capabilities in the general domain, they perform very poorly for VQA tasks in specialized domains such as medical imaging. This study is a submission for ImageCLEFmed-MEDVQA-GI 2025 subtask 1 that explores the adaptation of the Florence2 model to answer medical visual questions on GI endoscopy images. We also evaluate the model performance using standard metrics like ROUGE, BLEU and METEOR. The code used in the experiments is publicly available at: \href{https://github.com/gauravparajuli/ImageCLEFmed-MEDVQA-GI-2025-Task1}{github.com/gauravparajuli/ImageCLEFmed-MEDVQA-GI-2025-Task1}.
\end{abstract}


\input{sections/1.introduction}

\input{sections/2.related_work}

\input{sections/3.task_overview_and_dataset}

\input{sections/4.methodology}

\input{sections/5.result_and_evaluations}

\input{sections/6.ablation_studies}

\input{sections/7.discussion}

\input{sections/8.conclusion}

\input{sections/acknowledgement}

\bibliographystyle{unsrt}  
\bibliography{references}

\end{document}

%% file: sections/1.introduction.tex
\section{Introduction}

The gastrointestinal (GI) tract consists of all major organs in the digestive system, including the esophagus, stomach, and intestines. The GI tract is vulnerable to a wide range of abnormal mucosal conditions, ranging from minor irritations to highly lethal diseases~\cite{gelberg2017pathophysiological,navarre2009diseases}. Globally, 4.8 million new cases are reported annually, resulting in approximately 3.4 million deaths per year~\cite{arnold2020global}, highlighting the high mortality rate associated with GI diseases. Endoscopy is a medical procedure in which a thin device called an endoscope is inserted directly into the body to view organs and other structures. This allows the doctor to diagnose and sometimes treat the condition without surgery. Although endoscopy is a gold-standard procedure, there is about a 20\% miss rate for polyp identification in the colon due to operator error~\cite{ahmad2019artificial, wang2019real}.

This highlights the need for AI system intervention in medical diagnostics~\cite{Chaichuk2025May}. Medical VQA (MEDVQA) combines natural language processing with image understanding to answer diagnostic questions related to medical images. Such an AI-enabled support system offers promise in helping healthcare professionals provide high-quality care on a large scale accurately and efficiently.

While many available multimodal LLMs perform decently out of the box in the general domain, these same LLMs struggle greatly with VQA tasks in specialized domains such as medical imaging. This raises the need for supervised fine-tuning on a specialized medical dataset. This study is a submission for ImageCLEFmed MEDVQA-GI 2025 subtask 1. It explores how a general-purpose model can be adapted for better performance in specialized domains like medical imaging via supervised fine-tuning.

%% file: sections/2.related_work.tex
\section{Related Work}
Numerous efforts have been made to advance the field of MEDVQA. Since 2018, the annual ImageCLEF MEDVQA benchmark has played a critical role in pushing the field forward. Previously, MEDVQA methods combined CNNs and BERT with attention-based fusion for radiology images~\cite{yan2019zhejiang,benabacha2021vqamed,benabacha2019vqamed}. In addition, the datasets lacked focus on GI endoscopy. However, with the introduction of the KVASIR~\cite{pogorelov2017kvasir} dataset and the recently introduced KVASIR-VQA~\cite{KvasirVQA}, this issue has been mitigated. Current datasets provide question–answer pairs on GI tract images~\cite{Gautam2025Jun}. With the emergence of transformer-based models that have demonstrated strong performance in various deep learning tasks, this work focuses on fine-tuning Florence2 using the KVASIR-VQA dataset to enable effective VQA on GI endoscopic images.

%% file: sections/3.task_overview_and_dataset.tex
\section{Task Overview and Dataset}

We participated in subtask 1 (VQA) of the ImageCLEFmed 2025~\cite{OverviewImageCLEF2025} challenge. The objective of this challenge was to develop a model capable of accurately interpreting and answering all the questions related to GI images.

For this task, we used the KVASIR-VQA dataset. The Kvasir-VQA dataset is a combination of HyperKvasir~\cite{HyperKvasir} and KvasirInstrument~\cite{jha2021kvasir}. It consists of 6500 images across five different categories:
\begin{table}[ht]
\centering
\caption{Image Category in Kvasir-VQA}
\begin{tabular}{lccc}
\hline
\textbf{Category} & \textbf{Count}  \\
\hline
Normal & 2500 \\
Polyps & 1000 \\
Esophagitis & 1000 \\
Ulcerative Colitis & 1000 \\
Instrument & 1000 \\
\hline
Total & 6500 \\
\hline
\end{tabular}
\label{tab:kvasirvqa-image-category}
\end{table}

Each of these 6500 images in KVASIR-VQA is accompanied by various question--answer pairs. There are a total of six question types in the dataset.
\newline
\newline
1. Yes/No \textit{Example: Does this image contain any finding?}
\newline
2. Single choice \textit{Example: What type of polyp is taken?}
\newline
3. Multiple choice\textit{ Example: Are there any anatomical landmarks in the images?}
\newline
4. Choice(Color) \textit{Example: What color is the abnormality?}
\newline
5. Location \textit{Example: Where in the image is the abnormality?}
\newline
6. Numerical counting \textit{Example: How many polyps are there in the image?}
\newline

Since about 58,800 question--answer pairs were available for the 6500 images, no attempt was made to augment the dataset by paraphrasing question--answer pairs.

%% file: sections/4.methodology.tex
\section{Methodology}
\subsection{Architecture Overview}
 In this study, we fine-tuned the Florence2 model from Microsoft on KVASIR-VQA for medical VQA. Florence2 uses prompt-based multitask learning, which enables it to perform a diverse set of tasks such as object detection, segmentation and image captioning without the need for task-specific heads~\cite{xiao2024florence}. The Florence2 architecture consists of (a) a vision encoder (DaViT) that converts images into visual token embeddings, (b) a text encoder that processes prompt-style questions, (c) a multimodal transformer encoder--decoder that fuses image and text tokens and (d) a generative output that follows autoregressive decoding schemes like other LLMs.

\subsection{Training Details}
No augmentations were made to the training dataset provided by the organizer. The dataset was divided (using a seed value) into training and validation sets in a 9:1 ratio. We used LoRA (Low-Rank Adaptation) adapters for fine-tuning. Training was carried out for 5 epochs (early stopping patience set to 2 epochs) with evaluation at the end of each epoch on an NVIDIA RTX A4000 16GB GPU. We used Weights and Biases (wandb) to track training progress.

\subsubsection{Hyperparameters Search}
In order to determine the optimal hyperparameters, we used Optuna with Bayesian optimization to perform a hyperparameter search. We performed 100 search trials (one epoch each) using only 2.5\% of the actual dataset. Across all trials, the seed was fixed for reproducibility. Hyperparameter search ranges for the trials were: (1) learning rate: [1e-6, 1e-4], (2) batch size per device: [2, 4], (3) gradient accumulation steps: [1, 2, 4], (4) weight decay: [0.0, 0.1], (5) LoRA rank: [4, 8, 16], (6) LoRA alpha: [8, 16, 32] and (7) LoRA dropout: [0.0, 0.3]. After 100 search trials, the best configuration found was: (a) learning rate: 9.59e-5, (b) batch size per device: 2, (c) gradient accumulation steps: 2, (d) weight decay: 0.071, (e) LoRA rank: 16, (f) LoRA alpha: 32, (g) LoRA dropout: 0.05478.

However, the initial training with the above hyperparameters on the entire training set led to gradient explosion. Therefore, the learning rate obtained from the search was scaled down by a factor of four (referred to as the base learning rate from now on). Finally, to reduce the noise in the training loss curve, the effective batch size was increased to 64 via gradient accumulation and the base learning rate was scaled as:
\[
\sqrt{\frac{\text{new\_batch\_size}}{\text{old\_batch\_size}}}
\]
\\
Rest of the hyperparameters were kept as per the hyperparameters search result.

%% file: sections/5.result_and_evaluations.tex
\section{Results and Evaluation}

\begin{figure}[t!]
    \centering

    \begin{subfigure}[t]{0.32\textwidth}
        \centering
        \includegraphics[width=\linewidth]{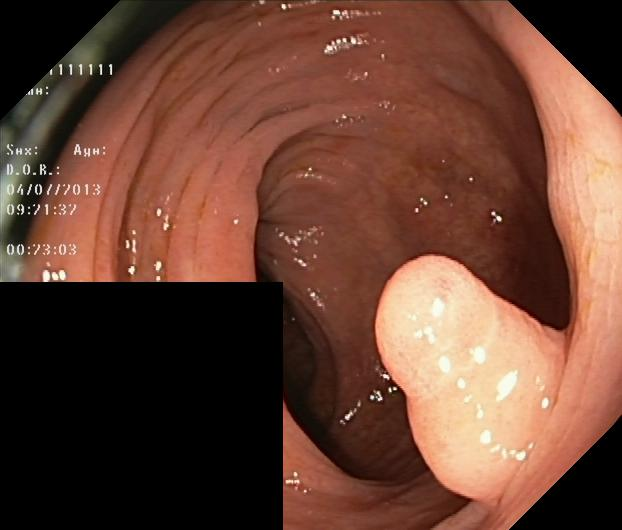}
        \scriptsize
        \vspace{0.5em}
        \begin{minipage}[t][2.8cm][t]{\linewidth}  
            \begin{flushleft}
                \textbf{Question:} Where in the image is the abnormality?\\
                \textbf{Prediction:} center; center-right; lower-right\\
                \textbf{Ground Truth:} center; center-right; lower-center; lower-right
            \end{flushleft}
            \vfill
        \caption{VQA on GI Endoscopy Image (Location Question)}
        \label{fig:location_vqa}
        \end{minipage}
    \end{subfigure}
    \hfill
    \begin{subfigure}[t]{0.32\textwidth}
        \centering
        \includegraphics[width=\linewidth]{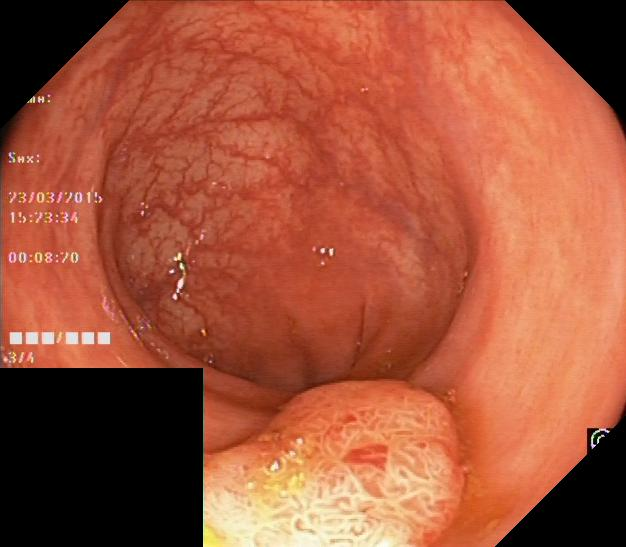}
        \scriptsize
        \vspace{0.5em}
        \begin{minipage}[t][2.8cm][t]{\linewidth}
            \begin{flushleft}
                \textbf{Question:} Are there any abnormalities in the image? Check all that are present. \\
                \textbf{Prediction:} polyp\\
                \textbf{Ground Truth:} polyp
            \end{flushleft}
            \vfill
            \caption{VQA on GI Endoscopy Image (Multiple Choice Question)}
            \label{fig:multiple_choice_vqa}
        \end{minipage}
    \end{subfigure}
    \hfill
    \begin{subfigure}[t]{0.32\textwidth}
        \centering
        \includegraphics[width=\linewidth]{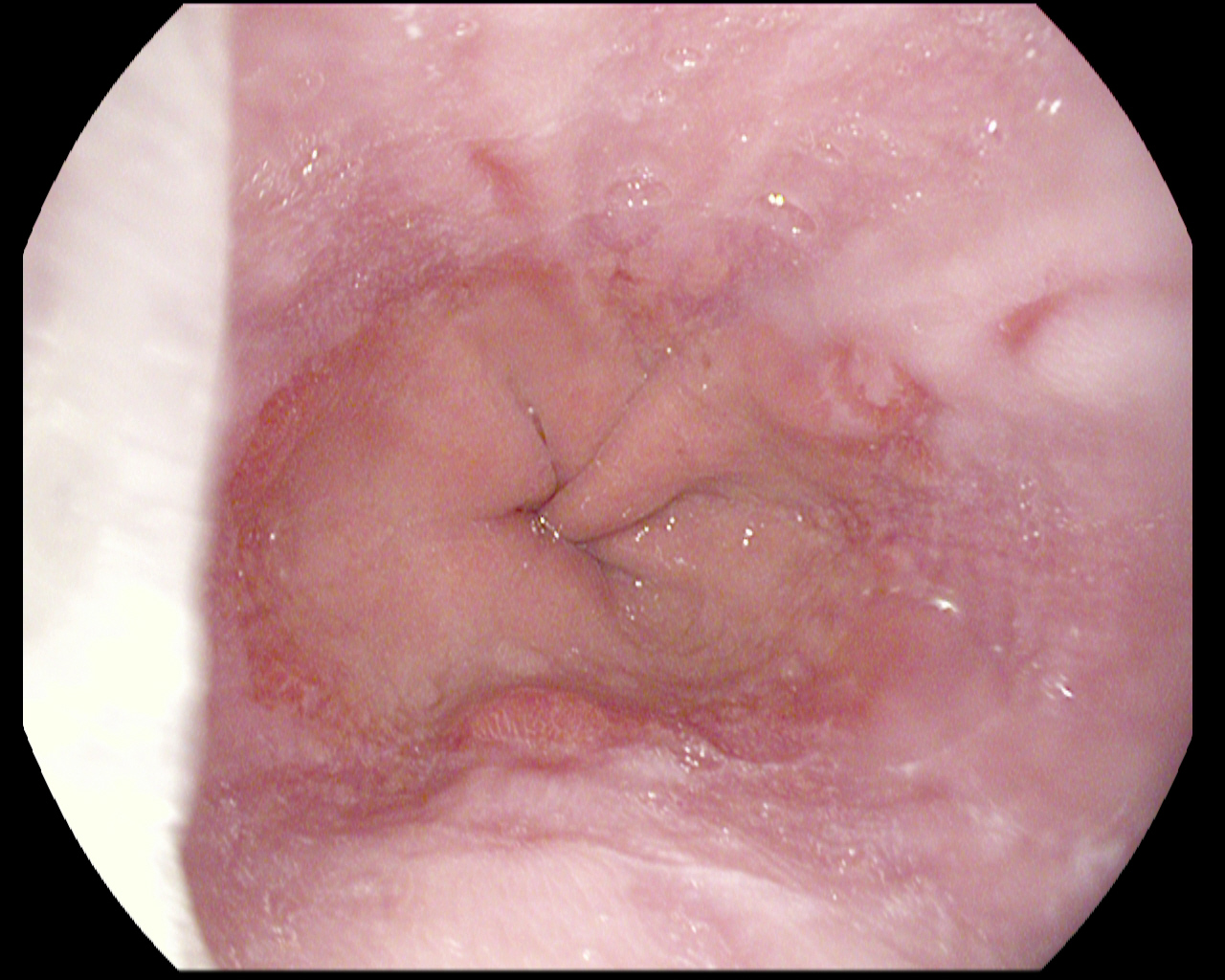}
        \scriptsize
        \vspace{0.5em}
        \begin{minipage}[t][2.8cm][t]{\linewidth}
            \begin{flushleft}
                \textbf{Question:} How many polyps are in the image? \\
                \textbf{Prediction:} 0\\
                \textbf{Ground Truth:} 0
            \end{flushleft}
            \vfill
            \caption{VQA on GI Endoscopy Image (Numerical Counting Question)}
            \label{fig:numerical_counting_vqa}
        \end{minipage}
    \end{subfigure}

    \vspace{1em} 

    \begin{subfigure}[t]{0.32\textwidth}
        \centering
        \includegraphics[width=\linewidth]{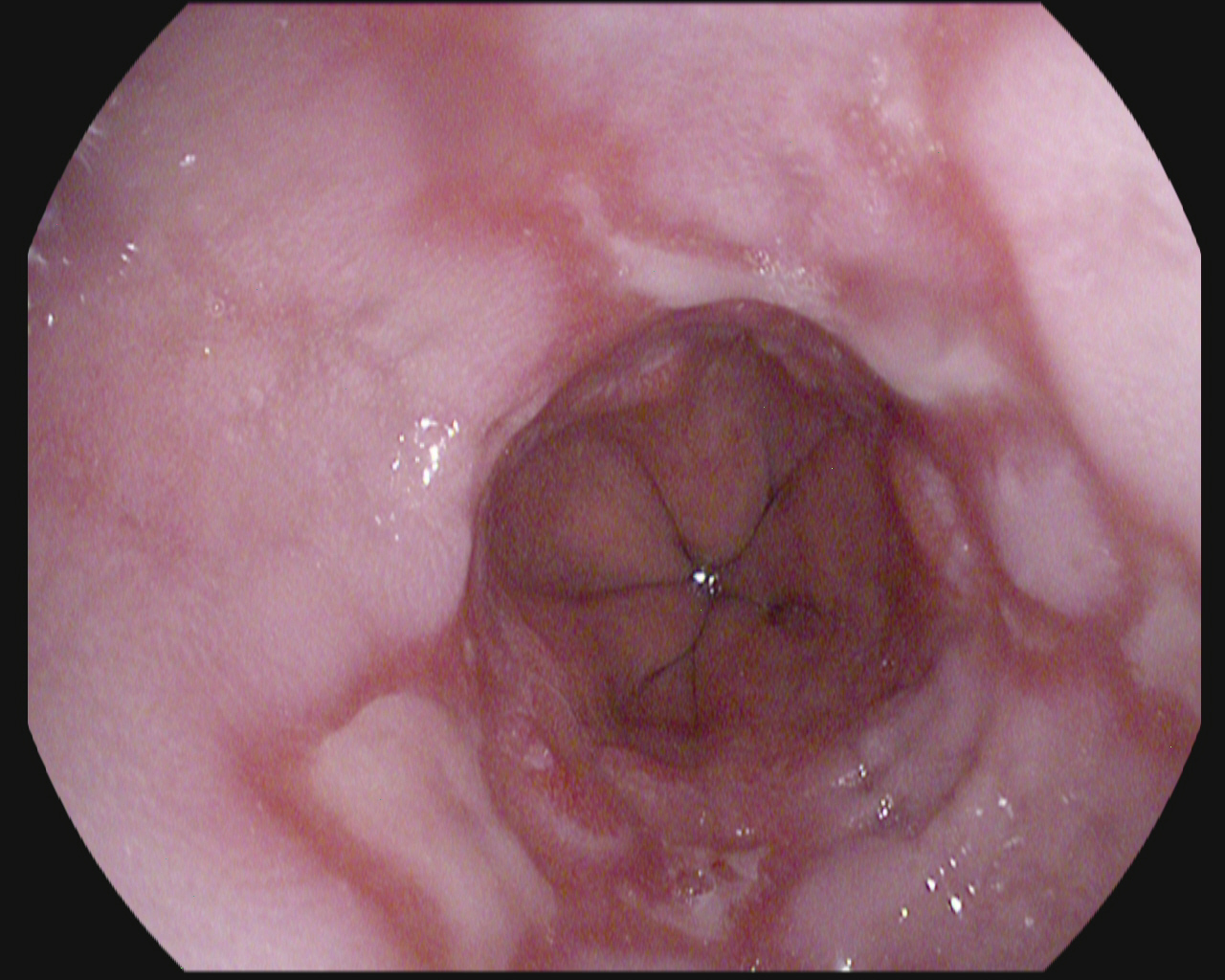}
        \scriptsize
        \vspace{0.5em}
        \begin{minipage}[t][2.5cm][t]{\linewidth}
            \begin{flushleft}
                \textbf{Question:} What color is the abnormality? If more than one separate with ; \\
                \textbf{Prediction:} pink; red; white\\
                \textbf{Ground Truth:} pink; red; white
            \end{flushleft}
            \vfill
            \caption{VQA on GI Endoscopy Image (Choice Color Question)}
            \label{fig:color_choice_vqa}
        \end{minipage}
    \end{subfigure}
    \hfill
    \begin{subfigure}[t]{0.32\textwidth}
        \centering
        \includegraphics[width=\linewidth]{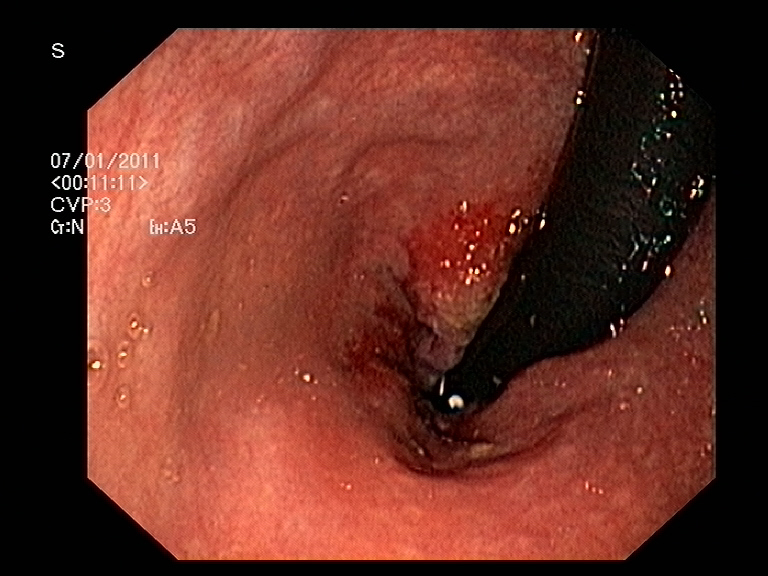}
        \scriptsize
        \vspace{0.5em}
        \begin{minipage}[t][2.5cm][t]{\linewidth}
            \begin{flushleft}
                \textbf{Question:} Is there a green/black box artefact? \\
                \textbf{Prediction:} no\\
                \textbf{Ground Truth:} no
            \end{flushleft}
            \vfill
            \caption{VQA on GI Endoscopy Image (Yes/No Question)}
            \label{fig:yes_no_vqa}
        \end{minipage}
    \end{subfigure}
    \hfill
    \begin{subfigure}[t]{0.32\textwidth}
        \centering
        \includegraphics[width=\linewidth]{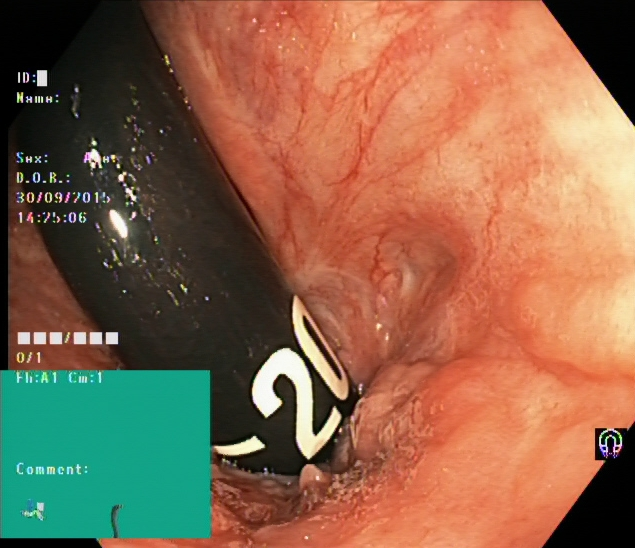}
        \scriptsize
        \vspace{0.5em}
        \begin{minipage}[t][2.5cm][t]{\linewidth}
            \begin{flushleft}
                \textbf{Question:} What type of procedure is the image taken from? \\
                \textbf{Prediction:} colonoscopy\\
                \textbf{Ground Truth:} colonoscopy
            \end{flushleft}
            \vfill
            \caption{VQA on GI Endoscopy Image (Single Choice Question)}
            \label{fig:single_choice_vqa}
        \end{minipage}
    \end{subfigure}

    \caption{VQA on GI Endoscopy Images for different question types}
    \label{fig:1}

    \centering
    \includegraphics[width=\linewidth]{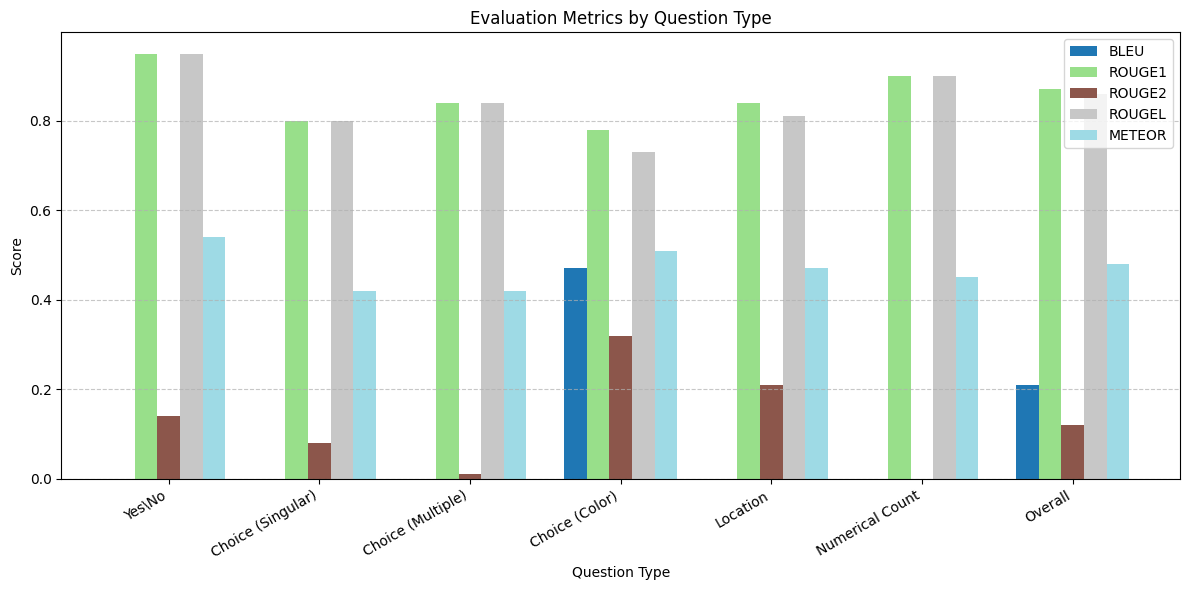}
    \caption{Evaluation Metrics across different question types}
    \label{fig:enter-label}
\end{figure}

The following table outlines the performance of our best model on both the public and private test sets of the organizing team.

\begin{table}[ht]
\centering
\caption{Evaluation Metrics for Test Set}
\begin{tabular}{lccccc}
\hline
\textbf{Test Set} & \textbf{BLEU} & \textbf{ROUGE1} & \textbf{ROUGE2} & \textbf{ROUGEL} & \textbf{METEOR} \\
\hline
Public	& 0.21	& 0.87	& 0.12	& 0.86 & 0.48 \\
Private	& 0.18	& 0.91	& 0.11	& 0.9 & 0.5 \\
\hline
\end{tabular}
\label{tab:results_metric}
\end{table}

The following table summarizes the performance of our best model across different question types.

\begin{table}[ht]
    \centering
    \caption{Evaluation Metrics for Test Set(across different question types)}
    \begin{tabular}{lccccc}
        \hline
         \textbf{Question Type} & \textbf{BLEU} & \textbf{ROUGE1} & \textbf{ROUGE2} & \textbf{ROUGEL} & \textbf{METEOR} \\
         \hline
         YesNo & 0.0 & 0.95 & 0.14 & 0.95 & 0.54 \\
         Choice (Singular) & 0.0 & 0.8 & 0.08 & 0.8 & 0.42 \\
         Choice (Multiple) & 0.0 & 0.84 & 0.01 & 0.84 & 0.42\\
         Choice (Color) & 0.47 & 0.78 & 0.32 & 0.73 & 0.51\\
         Location & 0.0 & 0.84 & 0.21 & 0.81 & 0.47\\
         Numerical Count & 0.0 & 0.9 & 0.0 & 0.9 & 0.45 \\
         \hline
         Overall & 0.21	& 0.87	& 0.12	& 0.86 & 0.48 \\
         \hline
    \end{tabular}
    \label{tab:results_metric_category_wise}
\end{table}
 
\clearpage

%% file: sections/6.ablation_studies.tex
\section{Ablation Studies}
To understand the impact of different fine-tuning strategies and the role of different components, we performed the following ablation studies using four variants of the Florence2 model.

\begin{table}[ht]
\centering
\caption{Ablation studies results for GI MEDVQA}
\begin{tabular}{lccccc}
\hline
\textbf{Model Variant} & \textbf{BLEU} & \textbf{ROUGE-1} &\textbf{ROUGE-2} & \textbf{ROUGE-L} & \textbf{METEOR} \\
\hline
florence2\_64\_FT & 0.2 & 0.85 & 0.11 & 0.85 & 0.47 \\
florence2\_64\_FT\_EF & 0.17 & 0.84 & 0.1 & 0.84 & 0.46\\
Florence2\_64\_r8 & 0.23 & 0.85 & 0.11 & 0.84 & 0.46 \\
\hline
Florence2\_64\_r16 & 0.21 & 0.87 & 0.12 & 0.86 & 0.48\\
\hline
\end{tabular}
\label{tab:ablation}
\end{table}

Please note that the model variant is in the format \texttt{baseModelName\_batchSize\_trainingStrategy}.

Model weights for the above variants are available at: \textit{https://www.hf.co/gauravparajuli/model\_variant\_name} \\

For the first variant, we froze the vision tower in the Florence2 architecture and proceeded with training. For the second variant, we froze the encoder portion of the Florence2 language model in addition to the vision tower. For the third and fourth variants, we used LoRA adapters with rank=8, alpha=16 and rank=16, alpha=32 respectively.

Here we can clearly see that the LoRA-based variant with rank=16 outperforms all variants. The LoRA-based variant with rank=8 achieves the highest BLEU score, possibly due to a regularization effect from the reduced parameter count, but it slightly underperforms on all other metrics.

The second variant, in which both the encoder and the vision tower were frozen, performs slightly under the first variant, in which only the vision tower was frozen. This suggests that while fine-tuning the decoder alone can capture some useful adaptation, the encoder's contribution is crucial for optimal performance in the multimodal task.

In general, these results validate the effectiveness of LoRA-based fine-tuning on downstream tasks.

%% file: sections/7.discussion.tex
\section{Discussion and Comparison with Literature}
Our approach outshines the performance of the previous year's baseline, which had a ROUGE1 score of 0.6955. However, despite achieving a strong ROUGE score, our model performed worse than the previous year's baseline in terms of the BLEU score (0.21 vs. 0.3757). The previous baseline was trained on only 2000 images (a small subset). As the BLEU metric penalizes short candidates, it is plausible that our model, which possibly generated more diverse and concise answers due to greater data exposure, was disproportionately penalized.

Also, Kvasir-VQA contains several question types. For "Yes/no" question and single word answer question, higher ROUGE score is easier to achieve as the vocabulary is limited. However, the BLEU score in this case will be very low or zero if the prediction does not exactly match the single word ground truth. This is the reason why the BLEU score was zero for the majority of question types in Table \ref{tab:results_metric_category_wise}.

%% file: sections/8.conclusion.tex
\section{Conclusion and Future Work}
It is evident from this study that it is possible to train a robust MEDVQA system based on Kvasir-VQA. Notable future directions for this study include: \\
1. Expanding the current dataset by introducing more diverse images and question answer pairs. This might help in developing more robust and generalizable models. \\
2. Increase the performance of the model on the BLEU metric without sacrificing the performance on the ROUGE metric. This could involve exploring different decoding strategies and loss functions that encourage more grammatically correct sentences. \\
3. Extensive evaluation of the model in real world clinical settings to gauge potential for real world applications.

%% file: sections/acknowledgement.tex
\section{Acknowledgments}
We would like to thank the organizers from the SimulaMet Department of Holistic Learning who made this event possible. This work heavily relied on the KVASIR-VQA dataset, which was also compiled by SimulaMet, and we deeply appreciate their contribution. Additionally, we would like to thank the researchers from Microsoft for their Florence-2 model. Lastly, this work would not have been possible without Hugging Face. We heartily thank everyone who has contributed to the Transformers library and the PEFT library within the Hugging Face ecosystem.